\documentclass[prl, aps,twocolumn,showpacs,preprintnumbers,amsmath,amssymb]{revtex4}

\usepackage{graphicx}
\usepackage{dcolumn}
\usepackage{bm}
\usepackage{color}

\begin{document}


\title{Universal damping behavior of dipole oscillations of one-dimensional ultracold gases induced by quantum phase slips}
\author{Ippei Danshita}
\affiliation{
{Yukawa Institute for Theoretical Physics, Kyoto University, Kyoto 606-8502, Japan and}
\\
{Computational Condensed Matter Physics Laboratory, RIKEN, Wako, Saitama 351-0198, Japan}
}

\date{\today}

\begin{abstract}
We study superflow decay via quantum phase slips in trapped one-dimensional (1D) quantum gases through dipole oscillations induced by sudden displacement of the trapping potential.
We find the relation between the damping rate of the dipole oscillation $G$ and the phase-slip nucleation rate $\Gamma$ as $G\propto \Gamma/v$, where $v$ is the flow velocity. 
This relation allows us to show that damping of 1D Bose gases in optical lattices, which has been extensively studied in experiment, is due to quantum phase slips.
It is also found that the damping rate versus the flow velocity obeys the scaling formula for an impurity potential even in the absence of an explicit impurity.
We suggest that the damping rate at a finite temperature exhibits a universal crossover behavior upon changing the flow velocity.
\end{abstract}

\maketitle

Systems of optical lattices loaded with ultracold gases have offered unique opportunities for the studies of correlated many-body physics in one dimension (1D), owing to their extraordinary controllability and cleanness~\cite{bloch-08, cazalilla-11}. In typical experiments, one creates an array of 1D gases by focusing a strong 2D optical lattice to a 3D gas, and the transverse confinement is widely controllable so that thermal and quantum motions in the transverse direction are completely frozen. With such extremely 1D quantum gases, recent experiments have revealed intriguing phenomena that are in stark contrast to higher dimensions, including Tonks-Girardeau gases~\cite{paredes-04, kinoshita-04} and their non-ergodic dynamics~\cite{kinoshita-06}, a possible fermionic superfluid of the Fulde-Ferrell-Larkin-Ovchinnikov type~\cite{liao-10}, the pinning Mott transition by a shallow optical lattice~\cite{haller-10}, and strong suppression of superfluid transport~\cite{haller-10, stoeferle-04, fertig-05, mun-07, gadway-11}.

Transport of trapped ultracold gases through periodic~\cite{haller-10, stoeferle-04, fertig-05, burger-01, mckay-08}, single-barrier~\cite{albiez-05}, or random potentials~\cite{gadway-11, pasienski-10} has been often investigated by suddenly displacing a parabolic trap to induce a dipole oscillation (DO) and observing its damping. As for the transport of 1D Bose gases, it has been found that the DO in the presence of an optical lattice~\cite{haller-10, stoeferle-04, fertig-05} or random potential~\cite{gadway-11} in the axial direction is significantly damped even in the superfluid state. 
Previous theoretical studies~\cite{anatoli-05, danshita-12} have suggested that this apparent contradiction, namely dissipative flow in a superfluid, can be interpreted as a consequence of phase slips (PS), in which thermal or quantum fluctuations cause the phase of the superfluid order parameter to unwind, leading to the dissipation of flow. 
This interpretation, if affirmative, could provide a unified view for superfluidity in 1D, given that the concept of PS is central also to the understanding of 1D superfluidity and superconductivity in several other condensed-matter systems, such as liquid $^4$He in 1D nanopores~\cite{taniguchi-10, eggel-11}, metallic nanowires~\cite{giordano-88, arutyunov-08}, and single-walled carbon nanotubes~\cite{kociak-01, tang-01, wang-12}.
Moreover, thanks to the flexible controllability of optical lattice systems, it would open up new possibilities for more thorough studies of PS.
However, relating explicitly the damping of DO to PS is highly non-trivial and has never been done thus far, because of difficulty in analyzing PS under a non-uniform trapping potential. 
Indeed, despite a number of previous numerical works on damped DO of 1D gases~\cite{anatoli-04, ruostekoski-05, rigol-05, ana-05, pupillo-06, danshita-09, montangero-09, okumura-10, anzi-11, huo-12}, the interpretation in terms of PS has never been corroborated.

In this Letter, we study the DO dynamics of trapped 1D superfluids in connection with PS. Through qualitative consideration on energy loss during the damping and exact numerical simulations with time-evolving block decimation (TEBD) method~\cite{vidal-04} at zero temperature, we find a parameter region where the damping rate $G$ and the PS nucleation rate $\Gamma$ satisfy the following simple relation as a function of the flow velocity $v$, 
\begin{eqnarray}
G(v) \propto \Gamma(v) / v.
\label{eq:dampdecay}
\end{eqnarray}
We emphasize that since the damping rate is a major experimental observable~\cite{haller-10, fertig-05, gadway-11}, the relation (\ref{eq:dampdecay}) allows for analyzing the PS nucleation rate in experiment.
Using this relation and TEBD, we show that the damping rate in 1D Bose gases in an optical lattice obeys the power-law formula derived from the nucleation rate of a quantum PS (QPS), thus numerically confirming the PS scenario. The exponent of the power-law is found to coincide with that for an impurity potential~\cite{kagan-00, buechler-01, cherny-11} rather than for a periodic potential~\cite{danshita-12}, although there is no explicit impurity. 
We also discuss the effects of finite temperatures to suggest a universal behavior of the damping rate in 1D superfluids that can be considered as a single-component Tomonaga-Luttinger (TL) liquid.

\begin{figure}[tb]
\includegraphics[scale=0.45]{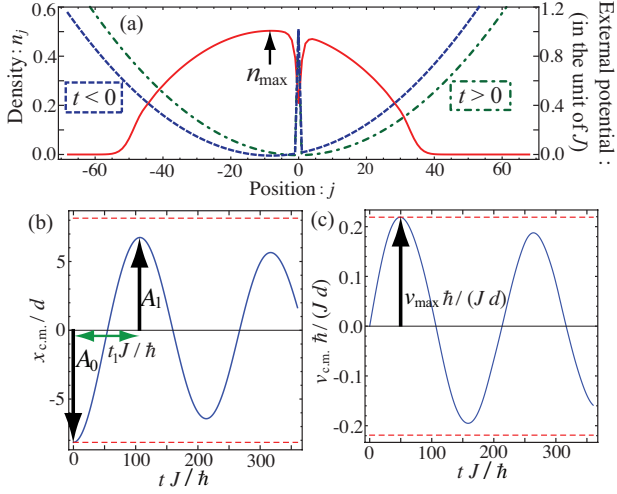}
\caption{\label{fig:HC_omnibus}
Numerical data for the Bose-Hubbard model Eq.~(\ref{eq:BHH}) in the hardcore limit ($U\rightarrow \infty$).
We set $N=31$, $V/J = -1.4$, $\Omega/J =0.00032$, $\lambda / J = 1$, and $x_0/d = 8$.
(a): The solid, dashed, and dashed-dotted lines represent the density distribution $n_j\equiv \langle \hat{n}_j \rangle$, the external potential at $t<0$, and that at $t>0$. 
(b) and (c): The time evolution of the center of mass position $x_{\rm c.m.}(t)$ and velocity $v_{\rm c.m.}(t)$.
}
\end{figure}

We describe 1D Bose gases of the total particle number $N$ by means of the following 1D Bose-Hubbard model,
\begin{eqnarray}
\hat{H} &=& -J\sum_{j}(\hat{b}^{\dagger}_j \hat{b}_{j+1} + {\rm h.c.}) +\frac{U}{2}\sum_{j}\hat{n}_j(\hat{n}_j -1)
\nonumber \\
&&\!\!\!\!\!\!\!\!\!\!
 + V\sum_{j}\hat{n}_j\hat{n}_{j+1} + \sum_{j}\left[ \Omega (j-X_c/d)^2 + \lambda \delta_{j,0}\right] \hat{n}_j,
\label{eq:BHH}
\end{eqnarray}
where $\hat{b}_j$ denotes the annihilation operator on the $j$th site and $\hat{n}_j = \hat{b}_j^{\dagger}\hat{b}_j$. 
$J$, $U$, and $V$ are the hopping energy, the onsite interaction, and the nearest-neighbor interaction, respectively. 
The last term in Eq.~(\ref{eq:BHH}) means the external potential that consists of a parabolic trapping potential and a single impurity,
%
%
where $\Omega$ is the trap curvature, $X_c$ the position of the trap center, $d$ the lattice spacing, and $\lambda$ the impurity strength.
While the system can be in an insulating state for certain values of the parameters, we hereafter assume the superfluid state because our main interest is in superflow decay via PS.

We use the TEBD method~\cite{vidal-04} to calculate the ground state and the exact quantum dynamics of the DO of Eq.~(\ref{eq:BHH}). TEBD is a quasi-exact numerical method for computing time evolution of a many-body wave function of 1D quantum lattice systems, and it can precisely describe quantum fluctuations causing QPS~\cite{schachenmayer-10, danshita-12, danshita-10}. To prepare an initial state for the real-time evolution, we first compute the ground state of Eq. (\ref{eq:BHH}) with the trap displacement $X_c = x_0$ $(>0)$ via imaginary time evolution. In Fig.~\ref{fig:HC_omnibus}(a), we show a typical external potential (dashed line) and density profile (solid line) of the initial state. With this initial state, we displace the trap to $X_c=0$ at $t=0$ as shown by the dashed-dotted line in Fig.~\ref{fig:HC_omnibus}(a) and compute the real-time evolution. In Figs.~\ref{fig:HC_omnibus}(b) and (c), we show an example of the time evolution of the center of mass (c.m.) position $x_{\rm c.m.}=N^{-1}d\sum_{j}j\langle \hat{n}_j \rangle$ and velocity $v_{\rm c.m.}=\dot{x}_{\rm c.m.}$, which exhibit a damped DO. We extract the damping rates using the formula,
%
$G = \ln\left(A_{\rm 0} / A_{\rm 1} \right)/t_1$.
%
%
%
As indicated in Fig.~\ref{fig:HC_omnibus}(b), $A_{\rm 0}$ and $A_{\rm 1}$ are the initial amplitude of $x_{\rm c.m.}/d$ and the amplitude after the half period $t_1$.

Before numerically verifying the relation (\ref{eq:dampdecay}) between $G$ and $\Gamma$, let us present a qualitative explanation that provides intuitive understanding of the relation. For this purpose, we express the oscillation-energy loss through the damping of the first half period in the following two ways. The first one is in terms of the lost potential energy,
%
$E_{\rm loss} = \frac{1}{2}M\omega^2(A_0^2-A_1^2)$,
where $\omega$ is the oscillation frequency and $M$ the total mass.
Assuming that the damping is so weak that $\delta \equiv 1-A_1/A_0 \ll 1$, or equivalently $Gt_1 \ll 1$, the energy loss can be rewritten as
\begin{eqnarray}
E_{\rm loss} \simeq M(\omega A_0)^2 \delta \simeq Mv_{\rm max}^2 Gt_1.
\label{eq:lost_pote}
\end{eqnarray}
Note that we used $\delta \simeq G t_1$ and $v_{\max} \simeq \omega A_{0}$ to derive Eq.~(\ref{eq:lost_pote}), where $v_{\max}$ is the maximum c.m.~velocity as indicated in Fig.~\ref{fig:HC_omnibus}(c). Thus, $E_{\rm loss}$ is expressed with $G$.
The other way is to use the Joule heat,
%
$E_{\rm loss} = P \times t_1$,
%
where $P=RI^2$ is the power, $R$ the resistance, $I \sim n_{\rm ave} v_{\rm max}$ the particle current, and $n_{\rm ave}$ the average density.
Assuming that the main source of the resistance is due to PS, the resistance can be related to the nucleation rate as $R = 2\pi \hbar \Gamma/I$~\cite{langer-67}. Using this relation, one obtains
\begin{eqnarray}
E_{\rm loss} \sim 2\pi \hbar n_{\rm ave}v_{\rm max} \Gamma t_1,
\label{eq:joule}
\end{eqnarray}
which connects the energy loss and the nucleation rate. 
Equating the right-hand side of Eq.~(\ref{eq:lost_pote}) with that of Eq.~(\ref{eq:joule}) leads to the relation, 
%
$G \sim 2\pi \hbar n_{\rm ave}/M \times \Gamma / v_{\rm max}$, 
%
which agrees with the relation~(\ref{eq:dampdecay}).

\begin{figure}[tb]
\includegraphics[scale=0.27]{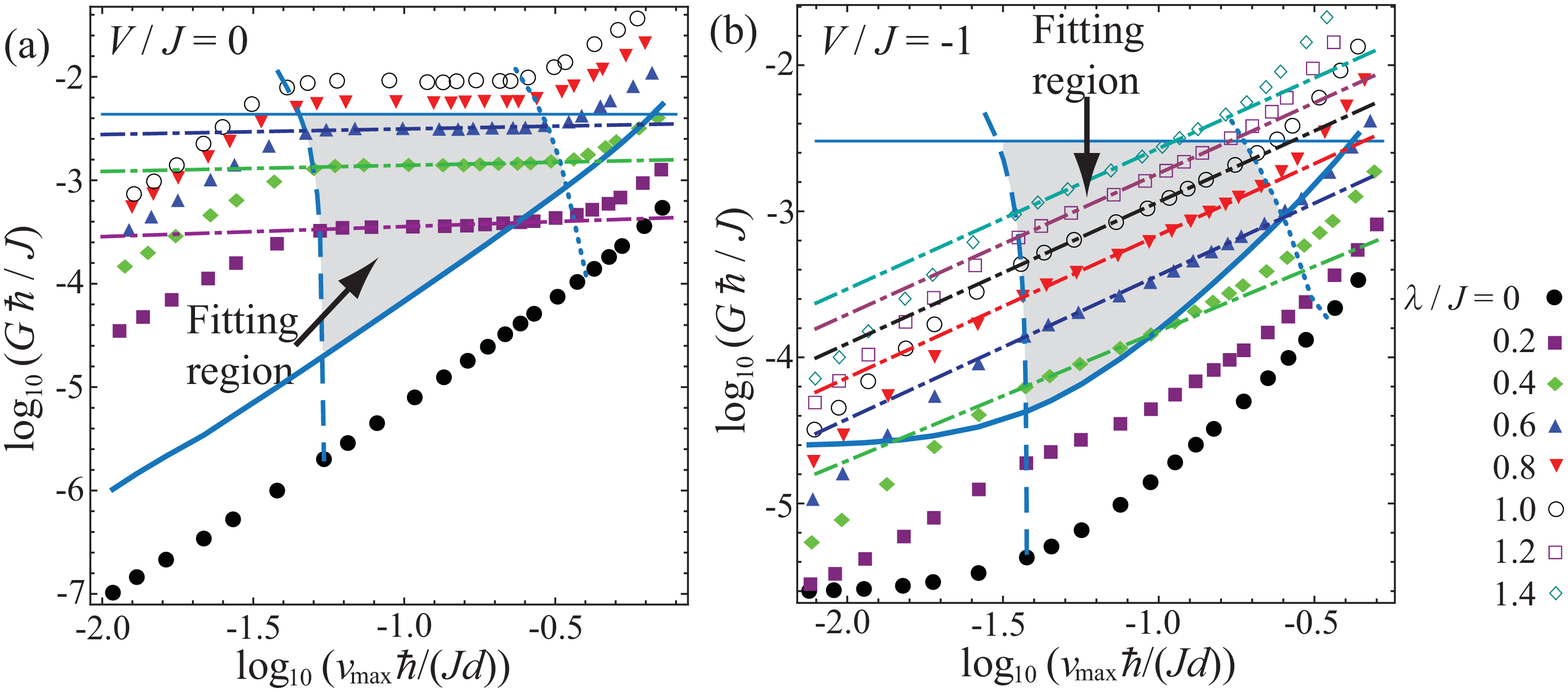}
\caption{\label{fig:HC_GvsV}
Damping rates $G$ versus the maximum flow velocity $v_{\rm max}$ for the hardcore Bose-Hubbard model with $N=31$ and several values of $\lambda/J$.
We set $(V/J, \Omega/J) = (0,0.001)$ (a) and $(-1,0.0005)$ (b). The thin solid lines represent $1/(4t_1)$ (in the unit of $J/\hbar$) at $\lambda = 0$ and $x_0 = d$. The thick solid lines represent $10G_0$. 
The dashed lines represent $v_{\rm max}$ at $x_0 = d$. The dotted lines represent $v_{\rm c}/5$. Numerical fitting with a function $\bar{G}(\bar{v}) = C \bar{v}^{\eta}$ is made for the data within the shaded region in order to extract the prefactor $C$ and the exponent $\eta$, and each dashed-dotted line is best fit to data for each $\lambda/J$. }
\end{figure}

Since the above explanation of the relation (\ref{eq:dampdecay}) is only qualitative, more quantitative corroboration is demanded. Hence, we analyze the DO in the hardcore boson limit ($U\rightarrow \infty$) using TEBD, in order to provide accurate numerical verification of the relation (\ref{eq:dampdecay}). In this limit, as long as $-2<V/J<0$, low energy physics of the system is known to be well described by the TL liquid model~\cite{kane-92}. Previous analytical studies have shown that the nucleation rate of a QPS in the TL liquid with a single impurity exhibits the following power-law behavior with respect to $v$ as $\Gamma_{\rm si} \propto v^{2K-1}$ for any $\lambda$ when $v\ll v_{\rm c}$~\cite{kagan-00, buechler-01}. Here $v_{\rm c}$ is the mean-field critical velocity and $K$ is the TL parameter~\cite{cazalilla-11}, the inverse of which, namely $1/K$, quantifies the strength of quantum fluctuations. 
To hold $K$ under control, we fix $N=31$, and adjust $\Omega/J$ depending on $V/J$ such that $n_{\rm max} \simeq 0.5$, where $n_{\rm max}$ is the maximum density [see Fig.~\ref{fig:HC_omnibus}(a)]. In such a situation, the analytical expression at half filling~\cite{cazalilla-11}, 
%
$K=\pi/[2\pi - 2\arccos \left({V \over 2J}\right)]$,
%
is approximately valid, and $K$ can be controlled by changing only $V/J$.

If the relation~(\ref{eq:dampdecay}) is correct,  the damping rate should obey $G \propto v^{2K-2}$.
To corroborate this, we plot in Figs.~\ref{fig:HC_GvsV}(a) and (b) $G$ versus $v_{\rm max}$ for $V/J = 0$ and $-1$, taking different values of $\lambda$. Note that we vary $x_0$ to control $v_{\rm max}$. As indicated by the shaded area in Fig.~\ref{fig:HC_GvsV}, we find the parameter region in which the damping rate safely obeys the power-law formula. This region is determined by the following four conditions: i) $Gt_1 < 1/4$, ii) $G>10G_0$, where $G_0$ is the damping rate at $\lambda=0$, iii) $x_0\geq d$, and iv) $v_{\rm max}<v_{\rm c}/5$~\cite{footnote-2}. We recall that the relation (\ref{eq:dampdecay}) is supposed to be valid when $Gt_1\ll 1$ and the source of the damping is mainly due to PS. While the first condition obviously corresponds to the former requirement, ii) and iii) stem from the latter. As for ii), we see from the black circles in Fig.~\ref{fig:HC_GvsV} that there is small but finite damping because of dephasing effects even in the absence of an impurity that induces QPS~\cite{rigol-05}. To distinguish the QPS from the dephasing, the damping at $\lambda > 0$ has to be much larger than that at $\lambda=0$, thus requiring ii). The condition iii) is necessary because otherwise a mismatch of the initial density and the displaced trap causes additional damping or revival of the DI that blurs the QPS effects. The last condition has to be satisfied to validate $\Gamma \propto v^{2K-1}$, as mentioned above.  These four conditions are indicated by the thin-solid, thick-solid, dashed, and dotted lines in Fig.~\ref{fig:HC_GvsV}. 

\begin{figure}[tb]
\includegraphics[scale=0.37]{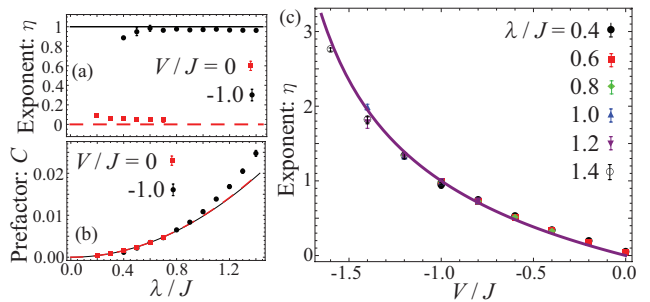}
\caption{\label{fig:HC_coef}
The exponent $\eta$ and the prefactor $C$ for the hardcore limit with $N=31$.
(a) and (b): $\eta$ and $C$ as functions of $\lambda/J$ for $(V/J, \Omega/J) = (0,0.001)$ (red circles) and $(-1,0.0005)$ (black squares). The red dashed and black solid lines in (a) represent $\eta = 2K-2$ for $V/J=0$ and $-1$.
The red dashed and black solid lines in (b) represent a parabolic function $f(\lambda/J)= a(\lambda/J)^2$ for $V/J=0$ and $-1$ with the constant $a$ determined such that the lines pass on the data points at $\lambda/J = 0.6$.
(c): $\eta$ versus $V/J$ for several values of $\lambda/J$. 
The solid line represents $\eta = 2K - 2$.
}
\end{figure}

The damping rates in the shaded area surrounded by the four lines indeed exhibit the power-law behavior.
By fitting a function $\bar{G}(\bar{v}) = C \bar{v}^{\eta}$ to the data in the area, we extract the exponent $\eta$ and the prefactor $C$, where $\bar{G}\equiv \hbar G/J$ and $\bar{v}\equiv \hbar v_{\rm max}/(Jd)$.
In Figs.~\ref{fig:HC_coef}(a) and (b), we plot $\eta$ and $C$ for $V/J=0$ and $-1$ as functions of $\lambda/J$.
When $\lambda/J$ increases, $\eta$ is almost constant and nearly equal to $2K-2$,  and $C$ quadratically increases for $\lambda < J$. 
This is consistent with the previous results that $\Gamma_{\rm si} \propto v^{2K-1}$ holds for any $\lambda$~\cite{kagan-00, buechler-01} and that $\Gamma_{\rm si} \propto \lambda^2$ for small $\lambda$~\cite{kagan-00}. 
In Figs.~\ref{fig:HC_coef}(c), we plot $\eta$ versus $V/J$ for different values of $\lambda/J$. 
There we see that the exponents agree very well with the expected value, i.e., $\eta = 2K -2$ that is represented by the solid line in Fig.~\ref{fig:HC_coef}(c).

\begin{figure}[t]
\includegraphics[scale=0.3]{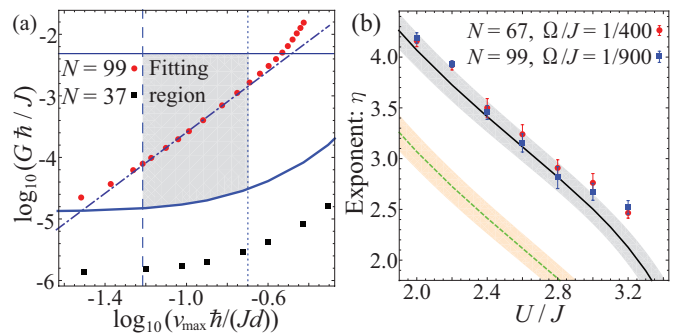}
\caption{\label{fig:soft}
The case of the softcore Bose-Hubbard model ($U<\infty$) with $V=0$ and $\lambda = 0$.
(a): $G$ versus $v_{\rm max}$ for $U/J = 3.2$ and $\Omega/J = 1/900$.
The thin solid, thick solid, dashed, and dotted lines represent $1/(4t_1)$ at $x_0 = d$, $10G_{\rm low}$, $v_{\rm max}$ at $x_0 = d$, and $v_{\rm max} = v_{\rm c}/8$. The dashed dotted line represents the best fit to the data inside the shaded region.
(b): $\eta$ versus $U/J$. The solid and dashed lines represent $2K-2$ and $2K-3$. 
The shaded regions mean the error bars of these lines originating from the errors in numerically evaluating $K$.
}
\end{figure}
Having corroborated the relation (\ref{eq:dampdecay}) in both qualitative and quantitative manners, we now consider the case of softcore bosons ($U<\infty$) without the nearest neighbor interaction and the impurity.  
This case is of direct relevance to the experiments of Refs.~\cite{stoeferle-04, fertig-05, haller-10, gadway-11}, where the damped DO of 1D Bose gases in optical lattices has been studied, and is of great importance for understanding whether the damping observed in the experiments is due to QPS. To address QPS effects, we choose $N$ and $\Omega$ such that $1 < n_{\rm max} < 2$. In this situation, there exist the regions of $n_j \simeq 1$, where the underlying lattice structure induces strong umklapp scattering leading to QPS. Since the nucleation rate of such a QPS obeys $\Gamma_{\rm prd} \propto v^{2K-2}$ when $v<v_{\rm c}$~\cite{danshita-12}, we naively speculate $G \propto v^{2K-3}$. We take $U<U_{\rm c}$ for the system to be safely in the superfluid state, where $U_{\rm c}\simeq 3.3 J$~\cite{ejima-11} is the Mott transition point.  

In Fig.~\ref{fig:soft}(a), we plot $G$ versus $v_{\rm max}$ for $N=99$, $\Omega/J = 1/900$, and $U/J=3.2$.
We again find the parameter region in which the damping rate exhibits a power-law behavior. While the conditions i) and iii) remain the same as the hardcore boson case, ii) and iv) are modified as ii$'$) $G>10G_{\rm low}$ and iv$'$) $v_{\rm max}<v_{\rm c}/8$, where $G_{\rm low}$ is the damping rate for $n_{\rm max} < 1$, e.g., $N=37$ is taken to be compared with the case shown in Fig.~\ref{fig:soft}(a). The condition ii$'$) guarantees that the main source of the damping is not the dephasing but the PS occurring at $n_j \simeq 1$. The reason for the latter modification will be discussed later.

By fitting a function $\bar{G}(\bar{v})=C\bar{v}^\eta$ to the data in the shaded area, we extract $\eta$ for $(N,\Omega/J)=(99,1/900)$ and $(67,1/400)$, and plot them as functions of $U/J$ in Fig.~\ref{fig:soft}(b). $\eta$ agrees with the value expected for an impurity potential, $2K-2$ (solid line), rather than that for a periodic potential, $2K-3$ (dashed line)~\cite{footnote-3}. This disagrees with the naive speculation mentioned above and is counter-intuitive in the sense that there is no explicit impurity in the system. However, it can be interpreted as follows.
When a Bose gas is trapped in a combined parabolic and periodic potential and $n_{\rm max} > 1$, there are two narrow regions where $n_j\simeq 1$. In these regions, the umklapp process is the most relevant, and the transport is strongly suppressed. Hence, the unit-filling regions move slower than the other parts of the gas, and act as impurities for the other parts. This interpretation also explains the requirement iv$'$) in the sense that the presence of the effective impurities may lower $v_{\rm c}$.

\begin{figure}[tb]
\includegraphics[scale=0.5]{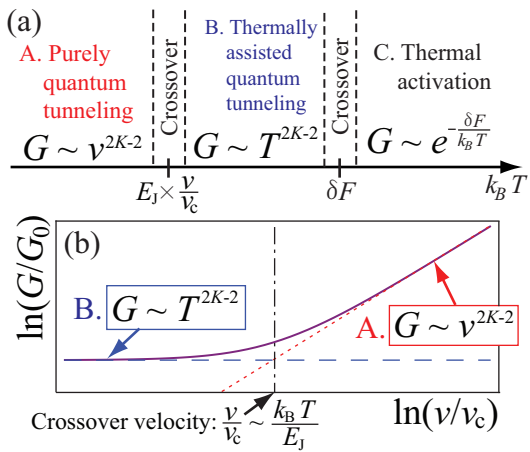}
\caption{\label{fig:finT}
(a): Schematic crossover diagram for the damping rate $G$ as a function of the temperature $T$.
(b): Sketch of the universal behavior in $G$ versus $v$ at a finite temperature.
}
\end{figure}
On the basis of the finding that the damping rate is related to the PS nucleation rate for an impurity potential both in the presence and the absence of an explicit impurity, we suggest that there are the following three distinct regimes with respect to the damping due to PS at finite temperatures, which are illustrated in Fig.~\ref{fig:finT}(a). (A) When the temperature is as low as $k_{\rm B}T\ll E_{\rm J}v/v_{\rm c}$, the PS is caused by pure quantum tunneling and the damping rate obeys $G \propto v^{2K-2}$, as discussed above. Here $E_{\rm J}=\hbar u/(\sqrt{2}d)$ is the Josephson plasma energy and $u$ is the sound velocity. (B) In the intermediate temperatures, $E_{\rm J}v/v_{\rm c}\ll k_{\rm B}T \ll \delta F$, the PS occurs due to the thermally assisted quantum tunneling~\cite{weiss-08} and $\Gamma \propto v T^{2K-2}$~\cite{kagan-00, buechler-01}, corresponding to $G \propto T^{2K-2}$, where $\delta F$ is the free energy barrier separating two neighboring winding-number states. (C) When $k_{\rm B}T \gg \delta F$, the thermal activation process becomes dominant~\cite{langer-67} and $G \propto e^{-\delta F/ (k_{\rm B}T)}$. 

As sketched in Fig.~\ref{fig:finT}(b), when $k_{\rm B} T \ll \delta F \sim E_{\rm J}$, the crossover between the regimes (A) and (B) can be induced by changing the flow velocity with a fixed temperature. The two regimes are separated by the crossover velocity that is $\sim v_{\rm c}\times k_{\rm B}T/E_{\rm J}$. Given that one can achieve $E_{\rm J}/k_{\rm B} \sim 30 \,{\rm nK}$~\cite{fertig-05} and $T\sim 4 {\rm nK}$~\cite{gadway-10} in current experiments, it is likely that the crossover can be observed. Since the main feature of this crossover is determined only by the TL parameter $K$ and the sound velocity $u$, we conjecture that it can be applied universally to the damping of DO in 1D quantum gases that can be effectively described as a single-component TL liquid. Examples include not only the two cases addressed above, but also paired or counterflow superfluid states of two-component Bose~\cite{anzi-11} or Fermi~\cite{okumura-10, huo-12} gases.

\label{sec:sum}
In conclusion, we have studied damped DO of trapped 1D quantum gases in the presence of an impurity potential or an optical lattice from a perspective of PS. 
We have connected the damping of DO and the PS nucleation through the relation (\ref{eq:dampdecay}).
Combining this relation with the TEBD simulations of the 1D softcore Bose-Hubbard model, we found that in certain parameter regions the damping rate algebraically grows with increasing the flow velocity as expected from the nucleation rate of a QPS.
This result strongly supports the interpretation that the strong suppression of superfluid transport observed in the experiments~\cite{stoeferle-04, fertig-05, haller-10, gadway-11} is due to QPS.
We also suggested a universal damping behavior at finite temperatures, which can be tested in future experiments.

\begin{acknowledgments}
The author thanks N. Prokof'ev for suggesting the relation (\ref{eq:dampdecay}) and A. Polkovnikov for critical reading of the manuscript and valuable comments.
Useful discussions with N. Bray-Ali, G. Pupillo, and D. Weiss are gratefully acknowledged.
The computation in this work was partially done using the RIKEN Cluster of Clusters facility.
\end{acknowledgments}



\begin{thebibliography}{99}
\bibitem{footnote-2}
Detailed mean-field analyses on the critical velocity of the hardcore Bose-Hubbard model in the presence of a single impurity will be presented elsewhere.
\bibitem{footnote-3}
Here $K$ is The TL parameter of the softcore Bose-Hubbard model at unit filling, which is numerically calculated via TEBD from the single-particle correlation function at equal time as done in Ref.~\cite{kuhner-00}.
\end{thebibliography}
\end{document}